\documentclass[12pt]{article}

\begin{document}
\title{Dynamical algebra and Dirac quantum modes in Taub-NUT background}

\author{Ion I. Cot\u{a}escu \thanks{E-mail:~~~cota@physics.uvt.ro}\\ 
{\small \it West University of Timi\c soara,}\\
       {\small \it V. P\^ arvan Ave. 4, RO-1900 Timi\c soara, Romania}
\and
Mihai Visinescu \thanks{E-mail:~~~mvisin@theor1.theory.nipne.ro}\\
{\small \it Department of Theoretical Physics,}\\
{\small \it National Institute for Physics and Nuclear Engineering,}\\
{\small \it P.O.Box M.G.-6, Magurele, Bucharest, Romania}}
\date{\today}

\maketitle

\begin{abstract}
The $SO(4,1)$ gauge-invariant theory of the Dirac fermions in the external 
field of the Kaluza-Klein monopole is investigated. It is shown that  the 
discrete quantum modes are governed by reducible representations of the 
$o(4)$ dynamical algebra generated by the components of the angular momentum 
operator and those of the Runge-Lenz operator of the Dirac theory in Taub-NUT 
background. The consequence is that there exist central and axial discrete 
modes whose spinors have no separated variables.

Pacs 04.62.+v

\end{abstract}

%\newpage

\section{Introduction}

In the relativistic quantum mechanics in the context of general relativity 
an important problem is how could be found the conserved operators 
which should be involved in the sets of commuting observables defining 
quantum modes. From this point of view examples of geometries giving rise 
to large sets of conserved quantities are useful for understanding the 
connection among symmetries and conservation laws in general relativity. 

One of the most interesting  geometries is that of the Euclidean Taub-NUT 
space since it has not only the usual isometries but also  admits a hidden 
symmetry of the Kepler type  if a cyclic variable is gotten rid of 
\cite{GM,GRFH}. 
Moreover in the Taub-NUT geometry there are four Killing-Yano tensors among 
them the first three are covariantly constant and realize the quaternionic 
algebra related to the hyper-K\" ahler structure of the Taub-NUT manifold. 
The last one, that exists by virtue of 
the metric being of type $D$, has a non-vanishing field strength.  
These Killing-Yano tensors represent a certain square root of three 
St\" ackel-Killing tensors  connected with
the components of an analogue to the Runge-Lenz vector of the 
Kepler type problem  since they give rise to constants of the geodesic 
motion (i.e. prime integrals) quadratic in 4-velocities 
\cite{GRFH,vH1,VV,MV}.

In other respects, the Taub-NUT metric is involved in many modern studies in 
physics. For example the Kaluza-Klein monopole was obtained by embedding the 
Taub-NUT gravitational instanton into five-dimensional Kaluza-Klein theory. 
On the other hand, in the long-distance limit, neglecting radiation, the 
relative motion of two monopoles is described by the geodesics of this 
space \cite{MAH}. The Dirac equation in the Kaluza-Klein monopole field was 
studied in the mid eighties \cite{DIRAC}. An attempt to take into 
account the Runge-Lenz vector of this problem was done in \cite{CH}.
We have continued this study showing 
that the Dirac equation is analytically solvable \cite{CV2} and determining 
the energy eigenspinors of the central modes. Moreover we derived all the 
conserved observables of this theory, including those associated with the 
hidden symmetries of the Taub-NUT geometry. Thus we obtained the Runge-Lenz 
vector-operator  of the Dirac theory, pointing out its specific properties 
\cite{CV3}.   

In the present paper we investigate  the consequences of the existence of 
the Runge-Lenz operator in the theory of the Dirac field in  
Taub-NUT background. We show that the dynamical algebras corresponding to 
different spectral domains are the same as in the scalar case but the 
representations of these algebras are different. Thus for each discrete 
energy level we obtain two irreducible representations of the $o(4)$ 
algebra determining different quantum modes. This is a new phenomenon due 
to the spin terms of the Runge-Lenz operator since in the scalar case  
for each energy level we have only one irreducible representation of the 
$o(4)$ algebra \cite{GRFH}. This conjecture offers us the opportunity to 
define new Dirac modes, called natural central or axial modes, the energy 
eigenspinors of which can be easily written using only algebraic properties.

In Sec.2 we briefly present the Killing vectors of the isometries of the 
Taub-NUT background and the Killing tensors corresponding to the hidden 
symmetries of this geometry, pointing out that these are related to the 
four specific Killing-Yano tensors. The general form of the conserved  
operators of the Dirac theory is discussed in the next section while in 
Sec.4 we write down the conserved operators arising from isometries or 
hidden symmetries, presenting their algebraic properties.  
Sec.5 is devoted to the study of the representations of the dynamical
algebra of the discrete energy levels. These can be completely investigated   
since the action of the Dirac operators can be calculated at the level of 
two-component Pauli spinors where the structure of the corresponding 
operators is simpler. This method helps us to define the natural discrete 
Dirac modes in the next section where, in addition, we completely determine 
the form of the energy eigenspinors of these modes in terms of the spinors 
of the simple modes defined in Appendix.    
We work in natural units with $\hbar=c=1$.

\section{The Taub-NUT geometry}

The background of the gauge-invariant five-dimensional theory of the Dirac 
fermions in the external field of the Kaluza-Klein monopole \cite{GPS} is 
the space of the Taub-NUT gravitational instanton with the time coordinate 
trivially added. Herein it is convenient to consider the static chart of 
Cartesian coordinates $x^{\mu}$, ($\mu, \nu,...=0,1,2,3,5$), with the line 
element
\begin{equation}\label{(met)} 
ds^{2}=g_{\mu\nu}dx^{\mu}dx^{\nu}=dt^{2}-\frac{1}{V}dl^{2}-V(dx^{5}+
A_{i}dx^{i})^{2}\,,
\end{equation}   
where $dl^{2}=(d\vec{x})^{2}=(dx^{1})^{2}+(dx^{2})^{2}+(dx^{3})^{2}$
is the usual Euclidean three-dimensional line element. This involves the 
Cartesian physical space coordinates $x^{i}$ ($i,j,...=1,2,3$) which 
cover the domain $D$. The other coordinates are the time, $x^{0}=t$, and 
the Cartesian Kaluza-Klein extra-coordinate, $x^{5}\in D_{5}$. Another 
chart suitable for applications is that of spherical coordinates, $x'=
(t,\,r,\,\theta,\,\phi,\,\chi)$, among them the first four are the time and 
the common spherical coordinates associated with $x^{i}$, ($i=1,2,3$), while 
$\chi+\phi=-\mu x^{5}$. The real number $\mu$ is the main parameter of the 
theory which enters in the expression of the function $1/V(r)=1+\mu/r$. 
The unique non-vanishing component of the vector potential in 
spherical coordinates is $A_{\phi}=\mu(1-\cos\theta)$. In Cartesian 
coordinates we have      
\begin{equation}
{\rm div}\vec{A}=0\,, \quad \vec{B}\,={\rm rot}\, 
\vec{A}=\mu\frac{\vec{x}}{r^3}\,.
\end{equation}
We note that the space domain $D$ is defined such that $1/V>0$.

The spacetime defined by (\ref{(met)}) has the symmetry given by the isometry 
group $G_{s}=SO(3)\otimes U_{5}(1)\otimes T_{t}(1)$ of rotations of the 
Cartesian space coordinates and $x^{5}$ and $t$ translations. The $U_{5}(1)$ 
symmetry is important since this eliminates the so called NUT singularity if 
$x^5$ has the period $4\pi\mu$. On the other hand, the $SO(3)$ isometry 
transformations involve all of the space coordinates, $x^i$ and  $x^5$, 
since these are given by a non-linear representation of the $SO(3)$ group. 
For this reason the corresponding Killing vectors, $k_{(i)}$ ($i=1,2,3$), 
have the components $k_{(i)}^j=-\varepsilon _{ijk}x^{k}$ and 
$k_{(i)}^5=-\varepsilon_{ijk}x^{j}A_{k}-\mu x^{i}/r$  
while the other two Killing vectors of the 
translation groups have usual constant components, 
$k_{(0)}^\mu=\delta^{\mu}_0$ and $k_{(5)}^{\mu}=\delta^{\mu}_5$.

The Taub-NUT geometry possesses four 
Killing-Yano tensors, $f^{(i)}$ ($i=1,2,3$) and $f^Y$, of valence 2  (which 
satisfy $f_{\mu\nu}=-f_{\nu\mu}$, and $f_{\mu\nu;\sigma}+f_{\mu\sigma;\nu}=0$). 
The first three, 
\begin{equation}\label{fi}
f^{(i)} 
= f^{(i)}_{\,\,{\hat \alpha}{\hat \beta}} {\hat e}^{\hat \alpha} \wedge 
{\hat e}^{\hat \beta}
=2 {\hat e}^5\wedge  {\hat e}^i +\varepsilon_{ijk} {\hat e}^j\wedge 
{\hat e}^k\,,
\end{equation}
are rather special since they are covariantly constant (with 
vanishing field strength). 
The fourth Killing-Yano tensor of the Taub-NUT space,  
\begin{equation}\label{fY}
f^{Y}=-\frac{x^i}{r}f^{(i)}+\frac{2x^i}{\mu V}\varepsilon_{ijk}\hat e^{j}\land
\hat e^{k}\,,
\end{equation}
is not covariantly constant having the non-vanishing components  of its 
field strength
\begin{equation}\label{stren}
f^{Y}_{\,r\theta;\phi}=\frac{2r^2}{\mu V}\sin\theta\,, 
\end{equation}
which are completely antisymmetric in $r,\,\theta$ and $\phi$.

The hidden symmetries of the Taub-NUT geometry are encapsulated in 
the non-trivial St\" ackel-Killing tensors $k_{(i)\mu\nu}, (i=1,2,3)$.
They can be expressed as symmetrized products of Killing-Yano tensors 
(\ref{fi}) and  (\ref{fY}) \cite{VV}:
\begin{equation}\label{kff}
k_{(i)\mu\nu} = -{\mu\over 4}(f^Y_{~\mu\lambda}f^{(i)\lambda}_{~~\nu} +
f^Y_{~\nu\lambda}f^{(i)\lambda}_{~~\mu})+{1\over 2\mu}(k_{(5)\mu}k_{(i)\nu} +
k_{(5)\nu}k_{(i)\mu}).
\end{equation}
In fact only the product of Killing-Yano tensors $f^{(i)}$ and $f^Y$ leads 
to non-trivial St\" ackel-Killing tensors, the last term in the r.h.s. 
of (\ref{kff}) being a simple product of Killing vectors. This term is 
usually added to write the Runge-Lenz vector in the standard form of the 
scalar (classical, Schr\" odinger or Klein-Gordon) theory 
\cite{GRFH,VV,CV1}.

\section{The operators of the Dirac theory}

The theory of the Dirac field in Cartesian charts of the Taub-NUT geometry 
takes the simplest form if one considers the local frames given by pentad 
fields, $e(x)$ and $\hat e(x)$, as defined in \cite{P}. Their components, 
have the usual orthonormalization 
properties and give the components of the metric tensor,  
$g_{\mu\nu}=\eta_{\hat\alpha \hat\beta}\hat 
e^{\hat\alpha}_{\mu}\hat e^{\hat\beta}_{\nu}$ and 
$g^{\mu\nu}=\eta^{\hat\alpha \hat\beta} e_{\hat\alpha}^{\mu}
e_{\hat\beta}^{\nu}$. In our notation \cite{CV2}, 
$\eta={\rm diag}(1,-1,-1,-1,-1)$ is the flat metric which raises or lowers   
the hated indices (ranging from 0 to 5). Its gauge group, $G(\eta)=SO(4,1)$,  
has as  universal covering group,    
$\tilde G(\eta)$, a subgroup of the group $SU(2,2)$ carried by the 
space of four-dimensional Dirac spinors. Therefore, the five 
matrices  $\tilde\gamma^{\hat\alpha}$, that must satisfy 
$\{ \tilde\gamma^{\hat\alpha},\, \tilde\gamma^{\hat\beta} \}
=2\eta^{\hat\alpha \hat\beta}$, 
can be defined in terms of the standard Dirac matrices \cite{DKK,TH} as 
$\tilde \gamma^{0}=\gamma^{0},\,      
\tilde \gamma^{i}=\gamma^{i}$, ($i=1,2,3$) and      
$\tilde \gamma^{5}=i\gamma^{5}$. These 
are self-adjoint with respect to the usual Dirac conjugation, i.e. 
$\overline{\tilde\gamma^{\hat\alpha}}=
\gamma^{0}(\tilde\gamma^{\hat\alpha})^{\dagger}\gamma^{0}
=\tilde\gamma^{\hat\alpha}$, 
and give the covariant basis generators of the group $\tilde G(\eta)$, 
denoted by  $S^{\hat\alpha \hat\beta}=i 
[\tilde\gamma^{\hat\alpha},\, \tilde\gamma^{\hat\beta}]/4$.

The  Dirac field $\psi$ of mass $M$, defined on the  space domain $D\times 
D_{5}$, has the gauge-invariant action \cite{DKK,DIRAC},      
\begin{equation}\label{(action)}
{\cal S}[\psi]=\int\, d^{5}x\sqrt{g}
\,\left\{\frac{i}{2}\left[\overline{\psi}\tilde\gamma^{\hat\mu}
\tilde\nabla_{\hat\mu}\psi-
(\overline{\tilde\nabla_{\hat\mu}\psi})\tilde\gamma^{\hat\mu}\psi
\right] -M\overline{\psi}\psi\right\} 
\end{equation}
where  $\tilde\nabla_{\hat\alpha}$ are the components of the  
spin covariant derivatives with local indices \cite{CV2},
\begin{equation}\label{sder}
\tilde \nabla_{i}=i\sqrt{V}P_{i}+\frac{i}{2}V\sqrt{V}\varepsilon_{ijk}
\Sigma_{j}^{*}B_{k}\,,\quad
\tilde \nabla_{5}=\frac{i}{\sqrt{V}}P_{5}-\frac{i}{2}V\sqrt{V}
\vec{\Sigma}^{*}\cdot\vec{B}\,.
\end{equation}
These depend on the momentum operators  
$P_{i}=-i(\partial_{i}-A_{i}\partial_{5})$ and $P_{5}=-i\partial_{5}$\,,
which in Taub-NUT geometry obey the commutation rules
$[P_{i},P_{j}]=i\varepsilon_{ijk}B_{k}P_{5}$ and
$[P_{i},P_{5}]=0$. 
The spin matrices which give the spin connection are defined by 
\begin{equation}\label{sigst}
\Sigma_{i}^{*}=S_{i}+\frac{1}{2}\gamma^{5}\gamma^{i} \,, \quad
S_{i}=\frac{1}{2}\varepsilon_{ijk}S^{jk}\,.
\end{equation}
The action (\ref{(action)}) leads to the Dirac equation, ${\cal D}\psi=M\psi$, 
given by the Dirac operator  \cite{DIRAC,CV2} 
\begin{eqnarray}
{\cal D}&=&i\tilde\gamma^{\hat\alpha}\tilde\nabla_{\hat\alpha}  
\nonumber=i\gamma^{0}\partial_{t}-{\cal D}_{s}\\ 
&=&i\gamma^{0}\partial_{t} - \sqrt{V}\vec{\gamma}\cdot\vec{P}
-\frac{i}{\sqrt{V}}\gamma^{5}P_{5}
-\frac{i}{2} V\sqrt{V}\gamma^{5}\vec{\Sigma}^{*}\cdot\vec{B}
\,,\label{(de)}
\end{eqnarray}
having the usual time-dependent term \cite{TH} and a {\em static} part, 
${\cal D}_{s}$.

In the standard representation of the Dirac matrices (with diagonal 
$\gamma^0$  \cite{TH}) the Hamiltonian operator is \cite{CV2}
$\tilde H=H+\gamma^{0}M$. The massless Hamiltonian,
\begin{equation}\label{HH}
H =\gamma^0{\cal D}_{s}=\left(
\begin{array}{cc}
0&V\pi^{*}\frac{\textstyle 1}{\textstyle \sqrt{V}}\\
\sqrt{V}\pi&0
\end{array}\right)\,,
\end{equation}
is expressed in terms of the operators $\pi=\,{\sigma}_{P}-iV^{-1}P_{5}$ and 
$\pi^{*}=\,{\sigma}_{P}+iV^{-1}P_{5}$ where $\sigma_P=\vec{\sigma}\cdot\vec{P}$ 
involves the Pauli matrices, $\sigma_i$. These Pauli operators give the space 
part of the  massless Klein-Gordon operator as,   
\begin{equation}
\Delta= V\,\pi^{*}\pi= V{\vec{P}\,}^{2}+\frac{1}{V}{P_{5}}^{2}\,.
\end{equation}

In what follows, we are interested especially by the form and the action of 
the {\em conserved} operators of the Dirac theory which, by definition, are 
the operators that {\em commute} with the Hamiltonian operator $\tilde H$. 
Since the mass term is simple and does not rise difficulties to keep 
the study of the operators as simple as possible we leave it aside.
We shall consider the massless case ($M=0$) and the 
Hamiltonian operator reduces to (\ref{HH}) without affecting the 
symmetries and conserved observables of the Dirac theory. The conserved 
operators depend on Pauli operators obeying several conditions requested 
by the above definition. Two types of Dirac operators are important for 
our further developments. The first one are the ${\cal Q}$-operators 
introduced in \cite{CV2} as  
\begin{equation}
{\cal Q}(X)=\left\{ H\,,\,\left(
\begin{array}{cc}
X&0\\
0&0
\end{array} 
\right) \right\}
=i\left[ Q_{0}\,,\,\left(
\begin{array}{cc}
X&0\\
0&0
\end{array} 
\right) \right] \,,
\end{equation}
where $Q_0=i{\cal D}_s=i\gamma^{0}H$. The linear mapping ${\cal Q}$ 
associates Dirac operators to the Pauli operators, $\pi,\,\pi^{*}$, 
$\sigma_{P}$, \,$\sigma_{L}=\,\vec{\sigma}\cdot\vec{L}$,  
$\sigma_{r}=\vec{\sigma}\cdot\vec{x}/r$, etc. \cite{DYON}.
The remarkable property of the ${\cal Q}$-operators is that if 
$[X,\,\Delta]=0$ then we can write
\begin{equation}
{\cal Q}(X){\cal Q}(Y)=H{\cal Q}(XY)\,,\quad
{\cal Q}(Y){\cal Q}(X)={\cal Q}(YX)H\,,
\end{equation}
for any other Pauli operator $Y$. Hereby it results that 
${\cal Q}(X)$ commutes with $H$ since $H={\cal Q}(1)$. Moreover, one can 
verify that if both the operators, $X$ and $Y$,  commute with $\Delta$ and 
commute or anticommute between themselves then ${\cal Q}(X)$ and 
${\cal Q}(Y)$ commute or anticommute  each other. An important 
property is that all of the 
${\cal Q}$-operators anticommute with $\gamma^0$. 
Another type of conserved operators are of the diagonal form 
$T={\rm diag}\,(T^{(1)},\,T^{(2)})$. Then the condition $[ T,\,H]=0$ requires 
the Pauli operators $T^{(1)}$ and $T^{(2)}$ to satisfy \cite{CV3}
\begin{equation}\label{T12}
T^{(2)}\sqrt{V}\pi=\sqrt{V} \pi T^{(1)}
\,,\quad V\pi^{*}\frac{1}{\sqrt{V}}T^{(2)}=
T^{(1)}V\pi^{*}\frac{1}{\sqrt{V}}\,.
\end{equation}
Hereby it results $[T^{(1)},\,\Delta]=0$ and the useful relations
\begin{equation}\label{TX}
 T{\cal Q}(X)={\cal Q}(T^{(1)}X)\,,\quad  
{\cal Q}(X) T={\cal Q}(XT^{(1)})\,,  
\end{equation}   
that can be used for deriving new algebraic properties.

\section{Symmetries and conserved observables}

A class of conserved operators are the generators of the 
operator-valued representations of the isometry group carried by the spaces 
of physical fields. In the scalar case, these are orbital operators defined 
up to the factor $-i$ as the Killing vectors fields associated to the 
isometries \cite{ES}. For the fields with spin the conserved generators get, 
in addition, specific spin terms determined by the form of the Killing vectors 
and the gauge fixing. Thus, in the case of the Dirac field (in four or five 
dimensions), for each Killing vector $k$ one can write the operator 
\cite{ES,CML} 
\begin{equation}\label{gen}
X_{k}=-ik^{\mu}\hat e_{\mu}^{\hat\alpha}\tilde\nabla_{\hat\alpha}+
\frac{1}{2}k_{\mu;\nu}e^{\mu}_{\hat\alpha}e^{\nu}_{\hat\beta}
S^{\hat\alpha \hat\beta}
\end{equation}
which {\em commutes} with the Dirac operator ${\cal D}$.

In Cartesian coordinates and our gauge fixing,  the  $U_{t}(1)$ generator is 
$-i\partial_{t}$, while the $U_{5}(1)$ one is $P_{5}$. In spherical 
coordinates it is convenient to replace $P_{5}$ with $Q=-\mu P_{5}=-i
\partial_{\chi}$.  The other three Killing vectors  give the 
$SO(3)$ generators which are the components of the whole angular momentum 
$\vec{\cal J} =\vec{L} + \vec{S}$  as in the flat spacetimes. The difference 
is that here the orbital angular momentum,   
\begin{equation}\label{(angmom)}
\vec{L}\,=\,\vec{x}\times\vec{P}-\mu\frac{\vec{x}}{r}P_{5}\,,
\end{equation} 
depends on $P_5$ since the $SO(3)$ isometries are non-linear transformations. 
The consequence is that the irreducible representations of the $o(3)$ 
algebra generated by $L_i$ are similar with the linear ones but with 
a supplementary restriction upon the angular quantum numbers \cite{CV1}.   
On the other hand, the operators ${\cal J}_i={\rm diag}(J_{i},\,J_i)$ have 
diagonal form where  $J_{i}=L_{i}+\sigma_{i}/2$ are just the angular momentum  
operators of the Pauli theory. For this reason ${\cal J}_{i}$ commute with 
${\cal D}$ and $\gamma^0$ and, therefore, they are conserved, commuting with 
$H$.   Moreover, they  satisfy the canonical commutation rules 
among themselves and with the components of all the other vector operators 
(e.g. coordinates,  momenta, spin, etc.).

However, there are other types of conserved operators directly related to 
the specific geometric objects of the Taub-NUT geometry, as St\" ackel-Killing 
and Killing-Yano tensors. The  Killing-Yano tensors give rise to 
conserved observables defined as Dirac-type operators of the form 
\begin{equation}\label{Qf}
Q_{f}=f_{\hat\alpha\hat\beta}
\tilde\gamma^{\hat\alpha}\tilde\nabla
^{\hat\beta}-\frac{1}{6}\gamma^{\mu}(x)\gamma^{\nu}(x)\gamma^{\lambda}(x)
f_{\mu\nu;\lambda}(x)
\end{equation}
where  
$f_{\hat\alpha\hat\beta}=f_{\mu\nu} e_{\hat\alpha}^{\mu} e_{\hat\beta}^{\nu}$
and $\gamma^{\mu}(x)=e^{\mu}_{\hat\alpha}\tilde\gamma^{\hat\alpha}$.  
According to an important result of Ref.\cite{CML}, these operators       
{\em anticommute} with the Dirac operator ${\cal D}$.

Starting with the first three Killing-Yano tensors,  from 
Eq.(\ref{Qf}), after some algebra, we obtain the Dirac-type operators  
\begin{equation}
Q_{i}=f^{i}_{\,\hat\alpha\hat\beta}\tilde\gamma^{\hat\alpha}\tilde\nabla
^{\hat\beta}={\cal Q}(\sigma_{i})\,,
\end{equation}
which anticommute with $Q_{0}$ and $\gamma^0$, commute with $H$ and obey
\begin{equation}
Q_{i}Q_{j}=\delta_{ij}H^{2}+i\varepsilon_{ijk}Q_{k}H\,,\quad 
[{\cal J}_{i}, Q_{j}]=i\varepsilon_{ijk}Q_{k}\,. 
\end{equation}
Hereby we find the $N=4$ superalgebra \cite{CV2}
\begin{equation}\label{QQH}
\{Q_{A},\,Q_{B}\}=2\delta_{AB}H^2\,, \quad A,B,...=0,1,2,3\,.
\end{equation}  
The corresponding Dirac-type operator of the fourth Killing-Yano tensor, 
$f^Y$, calculated  with the general rule (\ref{Qf}), is 
\begin{equation}\label{dy1}
Q^Y=-{\cal Q}(\sigma_r)+\frac{2i}{\mu\sqrt{V}}
\left(
\begin{array}{cc}
0&\lambda\\
-\lambda&0
\end{array}
\right)\,
\end{equation}
where $\lambda =\vec{\sigma}\cdot(\vec{x}\times\vec{P})+1=\sigma_{L}+1+
\mu\sigma_{r}P_{5}$ is the operator studied in \cite{JMP}. 
Other equivalent forms are given in \cite{CV3}.
Again one can verify that $Q^Y$ commutes with $H$ and anticommutes 
with $Q_{0}$ and $\gamma^0$.

Of a special interest is the Runge-Lenz operator of the Dirac theory 
associated to the Killing tensor $\vec{k}^{\mu\nu}$. This can be constructed
with the help of the conserved  Dirac-type operators generated by the 
Killing-Yano tensors. One defines first the vector operator 
$\vec{N}$ of components \cite{CV3} 
\begin{equation}
{N}_{i}=\frac{\mu}{4}\left\{ Q^{Y},\, Q_{i}\right\}-{\cal J}_{i}P_{5}
\end{equation}
which commutes with $H$  
since the operators $Q^Y$ and $Q_i$ are commuting with $H$. 
Consequently, $\vec{N}$ has diagonal form, its diagonal  blocks,   
$\vec{N}^{(1)}$ and $\vec{N}^{(2)}$, obeying the general conditions 
(\ref{T12}). One finds that the first block  which commutes with $\Delta$, 
\begin{equation}
\vec{N}^{(1)}=\vec{K}+\frac{\vec{\sigma}}{2}P_{5}\,, 
\end{equation}
contains not only the {\em orbital} Runge-Lenz operator, 
\begin{equation}\label{RLorb}   
\vec{K}=
\frac{1}{2}(\vec{P}\times \vec{L}-\vec{L}\times \vec{P})-
\frac{\mu}{2}\frac{\vec{x}}{r}\Delta +\mu\frac{\vec{x}}{r}{P_{5}}^{2}\,,
\end{equation}
but a spin term too. Furthermore, by taking into account that 
the components of $\vec{K}$ commute with $\Delta$ and satisfy 
\cite{GRFH}
\begin{eqnarray}
\left[ L_{i},\, K_{j} \right] &=& i \varepsilon_{ijk}\,K_{k}\,,\\ 
\left[ K_{i},\, K_{j} \right] &=& i \varepsilon_{ijk}L_{k}F^2\,, 
\quad F^2 ={P_5}^2-\Delta\,, 
\end{eqnarray}
one can calculate the algebraic properties  given in 
\cite{CV3} among them the commutation relations
\begin{equation}
\left[{N}_{i},\,{N}_{j}\right]= i\varepsilon_{ijk}{\cal J}_{k}
{\cal F}^2
+\frac{i}{2}\varepsilon_{ijk} Q_{k}H\,, \quad {\cal F}^2={P_5}^2-H^2\,,
\end{equation}
suggested us to define the components of the Runge-Lenz 
operator of the Dirac theory, $\vec{\cal K}$, as follows \cite{CV3}
\begin{equation}
{\cal K}_{i}={N}_{i}+ \frac{1}{2}H^{-1}({\cal F}-P_5) Q_i\,.
\end{equation}
Then one obtains the commutation relations 
\begin{eqnarray}
\left[{\cal K}_{i},\,H\right]=0\,,&\quad&
\left[{\cal K}_{i},\,{\cal J}_{j}\right]=i\varepsilon_{ijk}{\cal K}_{k}\,,\\
\left[{\cal K}_{i},\,P_5\right]=0\,,&\quad&
\left[{\cal K}_{i},\,Q_{j}\right]=i\varepsilon_{ijk}Q_{k}{\cal F}\,,
\end{eqnarray}
and
\begin{equation}
\left[{\cal K}_{i},\,{\cal K}_{j}\right]= i\varepsilon_{ijk}
{\cal J}_{k}{\cal F}^2\,.
\end{equation}
Since  there are no zero modes \cite{CV2} the 
operator $H$ is invertible such that our definition of the Runge-Lenz 
operator is correct.. It is worthy to note that ${\cal K}_{i}$ 
are diagonal commuting with $\gamma^0$ which means that they are also 
conserved in the massive case ($[{\cal K}_{i}, \tilde H]=0$).

\section{Dynamical algebra}   
 
The large collection of conserved observables we have obtained will help us 
to select many different complete sets of commuting observable which should 
define static quantum modes, with a given energy $E>0$. In other respects, 
since $P_{5}$ commutes with all  conserved observables, its eigenvalue, 
$\hat q$, will play the role of a general parameter. When one uses the 
operator $Q$ instead of $P_{5}$ then we have to consider as parameter its 
eigenvalue $q=-\mu \hat q$. 

In these conditions we can re-scale 
the Runge-Lenz operator in order to  recover the dynamical algebras 
$o(4)$, $o(3,1)$ or $e(3)$, corresponding to different spectral domains of 
the Kepler-type problems  \cite{GRFH}. If we define
\begin{equation}
{\cal R}_{i}=\left\{  
\begin{array}{lllll}
{\cal F}^{-1}{\cal K}_{i}&{\rm for}& \mu<0&{\rm and}&E<|\hat q|\\
{\cal K}_{i}&{\rm for}&{\rm any}~ \mu&{\rm and}&E=|\hat q|\\
i{\cal F}^{-1}{\cal K}_{i}&{\rm for}&{\rm any}~ \mu &{\rm and}& E>|\hat q|
\end{array}\right.
\end{equation}
then the operators ${\cal J}_i$ and ${\cal R}_i$ ($i=1,2,3$) will generate 
either a  representation of the $o(4)$ algebra for the discrete energy 
spectrum in the domain $E<|\hat q|$ or a representation of the $o(3,1)$ 
algebra for  continuous spectrum in the domain $E>|\hat q|$. The  dynamical 
algebra  $e(3)$  corresponds only to the ground energy of the continuous 
spectrum, $E=|\hat q|$. 

The manipulation of the dynamical algebras seems to be difficult in  view 
of the complicate structure of their generators but in fact the 
calculations can be done at the level of Pauli operators because of the 
special properties of the common eigenspinors of $H$ and $P_{5}$. These 
have the form  $U_{E,\hat q}=(u^{(1)}_{E,\hat q}, u^{(2)}_{E,\hat q})^{T}$ 
depending on two-component Pauli spinors  which  satisfy the equations    
\cite{CV2} 
\begin{eqnarray}
\Delta\, u_{E,\hat q}^{(1)}&=&E^{2}\,u_{E,\hat q}^{(1)}\,,\label{E1}\\
u^{(2)}_{E,\hat q}&=&E^{-1}\sqrt{V}\,\pi\,u^{(1)}_{E,\hat q}\,,\label{E2}
\end{eqnarray}
equivalent with the eigenvalue problem $(H-E)U_{E,\hat q}=0$. 
This means that $u_{E,\hat q}^{(1)}$ may be {\em any} solution of the scalar 
equation (\ref{E1}) which is just the static Klein-Gordon equation in 
Taub-NUT geometry. Moreover, it is clear that $u^{(2)}_{E,\hat q}$ is 
completely determined by Eq.(\ref{E2}) if the form of $u^{(1)}_{E,\hat q}$ 
is known. Since the Klein-Gordon equation is scalar, the form of the spinor  
$u^{(1)}_{E,\hat q}$ and, therefore, that of $U_{E,\hat q}$ is strongly 
dependent on the choice of the other observables included in the complete 
set of commuting operators which defines the quantum modes. Technically 
speaking, we have to start with the first Pauli spinor of the form  
$u^{(1)}_{E,\hat q}\sim \sum_{\sigma}f_{\sigma}\hat\xi_{\sigma}$ where 
$\hat\xi_{\sigma}$ 
are the usual Pauli eigenspinors of polarizations $\sigma=\pm 1/2$ while 
$f_{\sigma}$ are scalar solutions of Eq.(\ref{E1}). The next step is to 
precise this linear combination with the help of the other conserved 
operators of the complete set of commuting operators and to write 
$u^{(2)}_{E,\hat q}$ according to (\ref{E2}). The action of these  
operators can be easily calculated using the mentioned general properties 
(\ref{T12}) of the diagonal operators or the structure of the off-diagonal 
ones \cite{CV2}. For example, it is not difficult to show that the action 
of the Runge-Lenz operator upon the energy eigenspinors is  
\begin{equation}\label{cue}
\vec{\cal K}U_{E,\hat q}=\left(
\begin{array}{r}
\vec{K}'\,u^{(1)}_{E,\hat q}\\
E^{-1}\sqrt{V}\, \pi\vec{K}'\,u^{(1)}_{E,\hat q}
\end{array}\right)\,,\quad
\vec{K}'=\vec{K}+F\,\frac{\vec{\sigma}}{2}\,,
\end{equation}
which means that this reduces to the action of the new operator $\vec{K}'$ 
upon the first Pauli spinor. We say that $\vec{K}'$ is the  Pauli operator
{\em associated} with $\vec{\cal K}$. In fact, because of the central role 
played here by $u^{(1)}_{E,\hat q}$,  this property is general, each 
conserved observables having its own associated Pauli operator. Thus any  
problem of the Dirac theory in Taub-NUT background can be rewritten in 
terms of  Pauli operators acting upon the first two-component spinors of 
the spinors $U_{E,\hat q}$.

In order to illustrate how works this mechanism, let us study the 
representations of the $o(4)$ dynamical algebra of the discrete quantum modes 
of the Dirac field. As mentioned, these arise in the domain $E<|\hat q|$, 
only when $\mu<0$, and have the same energy levels as the scalar Klein-Gordon 
(or Schr\" odinger) \cite{CV2} equation,
\begin{equation}\label{(een)}
E_{n}^{2}=\frac{2}{\mu^2}\left[n\sqrt{n^{2}- q^{2}}-(n^{2}
- q^{2})
\right]\,.
\end{equation}        
These are labeled  only by the principal quantum number $n$ which takes 
all the integer values larger than $|q|$ \cite{CV2}. 
The quantization rule of the scalar modes is related to the irreducible 
unitary finite-dimensional representations of the $o(4)$ dynamical algebra 
generated by $L_{i}$ and the components of the re-scaled Runge-Lenz vector, 
$R_{i}=F^{-1}K_{i}$. Since the second Casimir operator of this algebra is 
$C_{2}=\vec{L}\cdot \vec{R}=0$, the first one, 
$C_{1}=\vec{L}^{2}+\vec{R}^{2}$,
can take only the eigenvalues $c_{1} =n^2 - 1$ (with $n>|q|$) which give 
the energy levels (\ref{(een)}).  In terms of $su(2)$ weights these 
representations of $o(4)\sim su(2)\times su(2)$ are denoted by 
$(\frac{n-1}{2}, \frac{n-1}{2})$.         

In the Dirac case the dynamical algebra is the same but its representations 
are generated by ${\cal J}_{i}$ and ${\cal R}_{i}={\cal F}^{-1}{\cal K}_{i}$.  
According  to Eq.(\ref{cue}), these representations are  {\em equivalent} 
with those generated by the associated Pauli operators $J_{i}$ and 
$R_{i}'=F^{-1}K_{i}'$ acting upon the first Pauli spinor. However, since 
$J_{i}=L_{i}+\sigma_{i}/2$ and $R_{i}'=R_{i}+\sigma_{i}/2$, we draw the 
conclusion that the Dirac discrete modes are governed by the reducible 
representation
\begin{equation}\label{desc}
\left(\frac{n-1}{2},\frac{n-1}{2}\right)\otimes\left(\frac{1}{2},0\right)=
\left(\frac{n}{2},\frac{n-1}{2}\right)\oplus \left(\frac{n}{2}-1,
\frac{n-1}{2}\right)\,. 
\end{equation}
The Casimir operators,
$C_{1}'=\vec{J}^2+{\vec{R'}}^2$ and $C'_2=\vec{J}\cdot \vec{R}'$ take now 
the eigenvalues, $c_{1}'$ and $c_{2}'$, such that $c_{1}'-2c_{2}'=n^2 -1$ 
for both irreducible representations while 
$c_{2}'=(2n+1)/4$ for the 
representation  $(\frac{n}{2},\frac{n-1}{2})$ and 
$c_{2}'=-(2n-1)/4$ for the 
representation  $(\frac{n}{2}-1,\frac{n-1}{2})$. This suggests us to use the 
new Pauli operator 
\begin{equation}
C=2C_{2}'-1/2=
 \sigma_{R}+\sigma_{L}+1\,,\quad \sigma_{R}=\vec{\sigma}\cdot\vec{R}         
\end{equation}
in order to distinguish between the irreducible 
representations resulted from the decomposition (\ref{desc}). The advantage 
is that this has the simplest eigenvalues, $c=\pm n$.    

\section{Discrete quantum modes}

Since the energy levels $E_{n}$ are degenerated, we need to use complete sets 
of commuting 
operators for determining quantum modes. Fortunately the set of conserved 
observables is large enough to offer us many possibilities of choice. An 
appropriate option is to consider the {\em natural} modes involving only one 
irreducible representation of (\ref{desc}). The complete sets of commuting 
observables of these modes must include the operators $H$, $P_5$ and 
the new operator ${\cal C}=2\vec{\cal J}\cdot \vec{\cal R}-1/2$ associated 
with the Pauli  operator $C$ defined above. Then only two more operators 
we need for defining these Dirac quantum modes.

We say that the set 
$\{H,\,P_{5},\,{\cal C},\,\vec{\cal J}^{2},\,{\cal J}_{3}\}$  
defines the {\em natural central} modes. Its common 
eigenspinors, $U_{n,\hat q,c,j,m_{j}}$, correspond to the 
eigenvalues $E_{n}$, $\hat q$, $c$, $j(j+1)$ and $m_{j}$.         
Another possibility is to take the set  
$\{H,\,P_{5},\,{\cal C},\,{\cal R}_{3},\,{\cal J}_{3}\}$  
of the {\em natural axial} modes the common  eigenspinors of which, 
$U_{n,\hat q,c,m_{r},m_{j}}$, correspond to the eigenvalues 
$E_{n}$, $\hat q$, $c$, $m_{r}$ and $m_{j}$. On the other hand,  
our previous results \cite{CV1,CV2} lead to the conclusion that there are 
{\em simple} modes having eigenspinors with separated variables. These are 
central and axial simple modes  defined by the sets 
$\{H,\,P_{5},\,\vec{\cal J}^{2},\,{\cal J}_{3},\,{\cal Q}(\sigma_{L}+1)\}$  
and $\{H,\,P_{5},\,{\cal R}_{3},\,{\cal J}_{3},\, Q_3\}$ respectively, 
as it is shown in Appendix.  
Since neither ${\cal Q}(\sigma_{L}+1)$ nor $Q_3$ do not commute with 
${\cal C}$, it results  that the natural modes do not have eigenspinors with 
separated variables. Therefore these must be linear combinations of the  
spinors of simple modes. Then it is interesting to try to write down these 
linear combinations using only the algebraic method based on the 
relation between Dirac and Pauli operators.

According to this method, we observe the first Pauli spinors, 
$u^{(1)}_{n,\hat q,c,j,\,m_{j}}$, of the eigenspinors 
$U_{n,\,\hat q,\,c,\,j,\,m_{j}}$ of the natural central modes must be 
the eigenspinors of the set $\{\Delta,\, P_5 ,\,C,\,\vec{J}^2 ,\,J_3 \}$ 
corresponding to the eigenvalues $E_{n}^{2}$, $\hat q$, $c$, $j(j+1)$
and $m_{j}$. On the other hand, the superalgebra 
\begin{equation}
\left\{ \sigma_{R},\,\sigma_{L}+1\right\}=0\,,\quad
(\sigma_{R})^2=C_{1}-\vec{L}^2-\sigma_{L}\,,
\end{equation}
allows us to demonstrate that the first Pauli spinors of the eigenspinors of 
the simple central modes (\ref{sscc}) have the remarkable property  
\begin{equation}
\sigma_{R}\, u^{(1)\pm}_{n,\hat q,j,\,m_{j}}= \left[n^{2}-
\left(j+{\textstyle \frac{1}{2}}\right)^2\right]^{1/2} 
u^{(1)\mp}_{n,\hat q,j,\,m_{j}}\,. 
\end{equation}
Hereby it results that the Dirac eigenspinors of the natural 
central modes can be expressed as 
\begin{eqnarray}
U_{n,\hat q,c=\pm n,j,m_{j}}& =&
\frac{1}{\sqrt{2n}}\left[\pm\sqrt{n\pm \left(j+ 
{\textstyle \frac{1}{2}}\right)}\,
U_{n,\hat q, j,m_{j}}^{\pm}\right.\nonumber\\ 
&&~~~~~~~~\left.+\sqrt{n\mp \left(j+ {\textstyle \frac{1}{2}}\right)}\,
U_{n,\hat q, j,m_{j}}^{\mp}\right]\,. 
\end{eqnarray}

In the same way the eigenspinors of the natural axial modes can be expressed 
as linear combinations of the spinors of the simple axial modes  
presented in Appendix. As in previous case the calculations reduce to the 
eigenspinors of the  set of Pauli operators $\{\Delta,\, P_5,\,C,\,
R'_3,\,J_3\}$ associated to that of the Dirac operators of the natural 
axial modes. Using the identity $\{C,\,\sigma_{3}\}=2(R'_{3}+J_3)$ it is not 
difficult to show that the eigenspinors of the natural axial modes are
\begin{eqnarray}
U_{n,\hat q,c=\pm n,m_{r},m_{j}}& =&
\frac{1}{\sqrt{2n}}\left[\pm\sqrt{n\pm |m_r + m_j|}\,
U_{n,\hat q, m_{r},m_{j}}^{\pm}\right.\nonumber\\ 
&&~~~~~~~~\left.+\sqrt{n\mp |m_r +m_j|}\,
U_{n,\hat q, m_{r},m_{j}}^{\mp}\right]\,. 
\end{eqnarray}

Thus we see that in the Dirac theory the algebraic method offers us the 
mechanisms of constructing new quantum modes having no separated variables. 
We can say that this method  and  that of separation of variables complete 
each other, helping us to find many types of different quantum modes 
related among themselves.

\section{Concluding remarks}

The aim of this paper was to complete the previous studies of the Dirac 
equation in Taub-NUT space \cite{CV2,CV3}. In the case of the Dirac equation 
the existence of the Killing-Yano tensors in the Taub-NUT geometry allowed us
to construct  some Dirac-type operators the anticommutators of which 
generate operators that commute with the standard Dirac one. 
We are attempting to find non-trivial operators of the Dirac theory 
connected with the hidden symmetry of the  Taub-NUT space. These operators 
are constructed in Sec.4 and they represent the quantum analogue of the 
classical Runge-Lenz vector from the Kepler problem. 
Thus we get an example of  gauge-invariant theory of Dirac fermions in a 
geometry with high manifest or hidden symmetries for which we can write all 
the corresponding conserved operators. The main consequence is that we can 
explicitly use the dynamical algebra for defining new quantum modes.    

In the literature there is a detailed description of the classical 
geodesic motion for scalar particles and a quantum treatment through 
the Schr\" odinger equation in Taub-NUT space \cite{GM,GRFH,CV1}. 
On the other hand, the pseudo-classical limit of the Dirac theory of a 
spin$-{1\over 2}$ fermion in curved spacetime is described by supersymmetric 
extension of the ordinary relativistic point particle \cite{BM}. In the 
pseudo-classical models the spin degrees of freedom are 
characterized in terms of anticommuting Grassmann variables. The 
constants of motion  related to the symmetries of the manifold 
generally contain spin-dependent parts \cite{GRH}. 
It is interesting that these specific spin contributions are similar to 
the spin terms of the conserved operators of the quantum theory. A 
notable example is the spin contribution to the Runge-Lenz vector which 
re-confirms the result from pseudo-classical approach \cite{VV}. Moreover, 
since both the pseudo-classical and the quantum theory are completely 
solvable, we have the opportunity to compare their physical meaning in 
all details.

\setcounter{equation}{0} \renewcommand{\theequation}
{A.\arabic{equation}}

\section*{Appendix A: The simple discrete modes}

The energy eigenspinors of the simple central modes have the form \cite{CV2}
\begin{equation}\label{sscc}
U^{\pm}_{n,\hat q,j,m_{j}}=\left(
\begin{array}{l}
u^{(1)\pm}_{n,\hat q,j,m_{j}}\\
u^{(2)\pm}_{n,\hat q,j,m_{j}}
\end{array}\right)
\sim\left(
\begin{array}{r}
f^{\pm}_{n,q,j}\,\Psi^{\pm}_{q,j,m_{j}}\\
h^{\pm} _{n,q,j}\,\Psi^{\pm}_{q,j,m_{j}}+
g^{\pm} _{n,q,j}\,\Psi^{\mp}_{q,j,m_{j}}
\end{array}\right)\,,
\end{equation}
where the radial functions $f^{\pm},\,g^{\pm}$ and $h^{\pm}$ depend only on 
$r$ while $\Psi^{\pm}$ are the two-component spherical spinors that  solve 
the common eigenvalue problems of the Pauli operators 
$Q=-\mu P_{5} ,\,{\vec{J}}^{2}$, $J_{3}$ and $\sigma_{L}+1$ for the 
eigenvalues $q=-\mu \hat q,\,j(j+1)$, $m_{j}$, and $\pm(j+{1\over 2})$, 
respectively \cite{CV2}. We observe that only the first Pauli spinor has 
separated variables. Its radial function, $f$, is a solution of the radial 
Klein-Gordon equation while the other two radial function arise from 
(\ref{E2}). The quantum numbers of these 
modes must satisfy the selection rule $|q|<j+\frac{1}{2}<n$.
Thus we find that $u^{(1)\pm}_{n,\hat q,j,\,m_{j}}$ is the common 
eigenspinor of the set $\{\Delta,\, P_5 ,\,\vec{J}^2 ,\,J_3 ,\, 
\sigma_{L}+1 \}$ corresponding to the eigenvalues 
$E_{n}^{2},\,\hat q,\, j(j+1),\, m_{j}$ and $\pm (j+\frac{1}{2})$ 
respectively. 
This means that (\ref{sscc}) is the common eigenspinor of the operators   
$\{H,\,P_{5},\,\vec{\cal J}^{2},\,{\cal J}_{3},\,{\cal Q}(\sigma_{L}+1)\}$  
whose eigenvalues are 
$E_{n},\,\hat q,\, j(j+1),\, m_{j}$ and $\pm E_{n}(j+\frac{1}{2})$.

The simple axial modes can be constructed in a similar way. We start with 
scalar common eigenfunctions $f_{n,\hat q, \tilde m, m}$ of the commuting 
operators $\Delta$, $P_5$, $R_3$ and $L_3$, corresponding to the eigenvalues 
$E_{n}^2$, $\hat q$, $\tilde m$ and $m$. These eigenfunction can be easily 
calculated in parabolic coordinates as the axial solutions of the 
Schr\" odinger equation \cite{CV1}, with the unique difference that the 
quantity $2E$ from the Schr\" odinger case must be replaced here by $E^2$. 
Furthermore, we construct the eigenspinors of the simple axial modes, 
$U^{\pm}_{n,\hat q, m_{r}, m_{j}}$, by choosing their first Pauli spinors as
\begin{equation}\label{uaa}
u^{(1)\pm}_{n,\hat q, m_{r}, m_{j}}\sim  f^{\pm}_{n,\hat q,\tilde m, m} 
\hat\xi_{\sigma=\pm1/2}
\end{equation}  
and calculating from (\ref{E2}) the second Pauli spinors that have 
no more separated variables. It is clear that 
(\ref{uaa}) is the common eigenspinor of the set of Pauli operators 
$\{\Delta,\,P_5,\,R'_{3},\,J_{3},\,\sigma_3\}$ associated 
with the Dirac ones from the set defining the simple axial modes. 
The eigenvalues of these operators are 
$E_{n}^2$, $\hat q$, $m_{r}=\tilde m+\sigma$, $m_{j}=m+\sigma$ and $\sigma$.   
Consequently,  $U^{\pm}_{n,\hat q, m_{r}, m_{j}}$ 
are the common eigenspinors of the set 
$\{H,\,P_{5},\,{\cal R}_{3},\,{\cal J}_{3},\, Q_3\}$ corresponding to the 
eigenvalues   
$E_{n}$, $\hat q$, $m_{r}$, $m_{j}$ and $E_{n}\sigma$.

\end{document}